\documentclass[aps, showpacs, showkeys,nofootinbib,floatfix]{revtex4}

\usepackage{amssymb}
\usepackage{amsmath}
\usepackage{graphicx}
\usepackage{hyperref}

\begin{document}

\title{Quasinormal modes of gravitational perturbation around a Schwarzschild black hole surrounded by quintessence}

\author{Yu Zhang}
\email{zhangyu128@student.dlut.edu.cn}
\author{Y.X.Gui\footnote{Corresponding author}}
\email{thphys@dlut.edu.cn}

\affiliation{Department of Physics, Dalian University of Technology,
Dalian, 116024, People's Republic of China}

\begin{abstract}
In this paper, the quasinormal modes of gravitational perturbation
around a Schwarzschild black hole surrounded by quintessence were
evaluated by using the third-order WKB approximation. Due to the
presence of quintessence, the gravitational wave damps more slowly.
\end{abstract}

\maketitle

\section {Introduction}
The quasinormal modes of a black hole present complex frequencies
that correspond to solutions of the perturbation equations, which
satisfy the boundary conditions appropriate for purely ingoing waves
at the horizon and purely outgoing waves at infinity.
Vishveshwara\cite{1} firstly pointed out the concept of QNMs in
calculations of the scattering of gravitational waves by a black
hole. The study of QNMs of black hole has a long
history\cite{2}-\cite{17}. The WKB\cite{18}-\cite{20} approach is a
semi-analysis method which was used by many scientists  to calculate
QNMs such as Schwarzschild\cite{21}, Reissener-Norstrom\cite{22},
Kerr\cite{23}, Kerr-Newman\cite{24}and so on . Recently, the
properties of quasinormal modes have been investigated in the
context of the AdS/CFT\cite{25}-\cite{31} correspondence and Loop
Quantum Gravity\cite{32}-\cite{34}.
\\ Recent measurements of the cosmic microwave background(CMB)\cite{35}
 combined with supernovae of type Ia\cite{36}, large scale structure (cosmic shear) \cite{37}
and galaxy cluster abundances\cite{38} show that 70\% of the
Universe is dominated by a mysterious dark energy which causes the
cosmic expansion to accelerate and 26\% cold dark matter, the
ordinary baryonic matter makes up only 4\%. There are several types
of models of dark energy, such as: the cosmological
constant\cite{39}, quintessence\cite{40}, phantom\cite{41},
k-essence\cite{42}, quintom\cite{43} models. For quintessence, the
equation of state is given by
\begin{math}p_{q}=w_{q}\rho_{q}\end{math} with
\begin{math}w_{q}\end{math} in the range of
$-1\leq w_{q} \leq-\frac{1}{3}$. Recently, Kiselev\cite{44}
considered  Einstein's field equations for a black hole surrounded
by the quintessential matter and obtained a new solution dependent
on the state parameter $ w_{q}$ of the quintessence. And Songbai
Chen et al\cite{45} has evaluated the quasinormal frequencies of
massless scalar field perturbation around the black hole which is
surrounded by quintessence. The result shows that due to the
presence of quintessence, the scalar field damps more rapidly. In
our paper, we investigate gravitational perturbation in this
situation.
\\  In this paper, the quasinormal modes of Schwarzschild black hole
surrounded by quintessence were evaluated by the third-order WKB
approximation.

\section{Quasinormal mode of gravitational perturbation around the black hole surrounded by quintessence}
The author\cite{44} gets the general forms of exact
spherically-symmetric solutions for the Einstein equations
describing black holes surrounded by the quintessence with the
energy momentum tensor, which satisfies the condition of the
additivity and linearity. The metric\cite{45} is given by
 \begin{equation}\label{eq:1}ds^{2}=(1-\frac{2M}{r}-\frac{c}{r^{3w_{q}+1}})dt^{2}
 -(1-\frac{2M}{r}-\frac{c}{r^{3w_{q}+1}})^{-1}dr^{2}-r^{2}(d\theta^{2}+sin\theta^{2}d\phi^{2}),\end{equation}
where M is the black hole mass, $w_{q}$ is the quintessential state
parameter, $c$ is the normalization factor related to
$\rho_{q}=-\frac{c}{2}\frac{3w_{q}}{r^{3(1+w_{q})}}$, and $\rho_{q}$
is the density of quitenssence.
\\  The study of black hole perturbations  was initiated by Regge
 and Wheeler\cite{46} for the \lq odd \rq parity types of harmonics  and was
 continued to the \lq even\rq parity by Zerilli\cite{47}.
 \\ Indicate the background metric with $g_{\mu\nu}$ and the
 perurbation in it with $h_{\mu\nu}$. The perturbation $h_{\mu\nu}$
 are supposed to be very small compared with $g_{\mu\nu}$. The
 $h_{\mu\nu}$ can be calculated from $g_{\mu\nu}$, and $R_{\mu\nu}+\delta
 R_{\mu\nu}$ from $g_{\mu\nu}+h_{\mu\nu}$. $\delta
 R_{\mu\nu}$ can be expressed in the form\cite{48}
 \begin{equation}\label{eq:2}
 \delta R_{\mu\nu}=-\delta \Gamma^{\beta}_{\mu\nu;\>\beta}+\delta
\Gamma^{\beta}_{\mu\beta;\>\nu},
\end{equation}
 where \begin{equation}\label{eq:3}
 \delta\Gamma^{\alpha}_{\beta\gamma}=\frac{1}{2}g^{\alpha\nu}(h_{\beta\nu;\>\gamma}+h_{\gamma\nu;\>\beta}-h_{\beta\gamma;\>\nu}),
 \end{equation}

The canonical form for the perturbations in the Regge-Wheeler gauge
is given as\cite{46}

\begin{equation}\label{eq:4}
 h_{\mu\nu}=
 \begin{array}{|cccc|}
  0&0&0&h_{0}(r)
 \\0&0&0&h_{1}(r)
 \\0&0&0&0
 \\sym&sym&0&0
 \end{array}\exp(-i\omega T)(\sin\theta/\partial\theta)P_{L}(\cos\theta)
 \end{equation}
Submit (\ref{eq:4}) to (\ref{eq:2}), we can get
\begin{equation}\label{5}
    (1-\frac{2M}{r}-\frac{c}{r^{3w_{q}+1}})^{-1}i\omega h_{0}+\frac{d}{dr}[(1-\frac{2M}{r}-\frac{c}{r^{3w_{q}+1}})h_{1}]=0\quad\quad\quad  from\quad \delta R_{23}=0
\end{equation}
\begin{equation}\label{6}
    (1-\frac{2M}{r}-\frac{c}{r^{3w_{q}+1}})^{-1}i\omega(\frac{dh_{0}}{dr}+i\omega h_{1}-\frac{2h_{0}}{r})+(l-1)(l+2)\frac{h_{1}}{r^{2}}=0 \quad\quad\ from\quad \delta R_{13}=0
\end{equation}
Defining
\begin{equation}\label{7}
    \Phi(r)=(1-\frac{2M}{r}-\frac{c}{r^{3w_{q}+1}})h_{1}/r
    \end{equation}
\\and
\begin{equation}\label{8}
    \frac{dr_{*}}{dr}=(1-\frac{2M}{r}-\frac{c}{r^{3w_{q}+1}})^{-1}
\end{equation}
and eliminate $h_{0}$, we then get
\begin{equation}\label{9}
    (\frac{d^{2}}{dr_{*}^{2}}+\omega^{2})\Phi(r)=V\Phi(r)
\end{equation}
where
\begin{equation}\label{10}
    V=(1-\frac{2M}{r}-\frac{c}{r^{3w_{q}+1}})(\frac{l(l+1)}{r^{2}}-\frac{6M}{r^{3}}-\frac{c(3w_{q}+3)}{r^{3w_{q}+3}})
\end{equation}
In this paper, we use the third-order WKB approximation method
devised by Schutz, Will\cite{18} and Iyer\cite{19}.The formula for
the complex quasinormal frequencies $\omega$ is
\begin{equation}\label{11}
    \omega^{2}=\left [V_{0}+(-2V''_{0})^{1/2}\Lambda\right ]-i(n+\frac{1}{2})(-2V''_{0})^{1/2}(1+\Omega)
\end{equation}
where
\begin{align*}
    \Lambda=\frac{1}{(-2V''_{0})^{1/2}}\left\{\frac{1}{8}\left(\frac{V^{(4)}_{0}}{V''_{0}}\right)(\frac{1}{4}+\alpha^{2})-\frac{1}{288}\left(\frac{V'''_{0}}{V''_{0}}\right)^{2}(7+60\alpha^{2})\right\}
\end{align*}
\begin{align}
   \Omega=&\,\frac{1}{(-2V''_{0})}\Big\{\frac{5}{6912}\left(\frac{V'''_{0}}{V''_{0}}\right)^{4}(77+188\alpha^{2})\nonumber\\
   &-\frac{1}{384}\left(\frac{V'''^{2}_{0}V^{(4)}_{0}}{V''^{3}_{0}}\right)(51+100\alpha^{2})+\frac{1}{2304}\left(\frac{V^{(4)}_{0}}{V''_{0}}\right)^{2}(67+68\alpha^{2})\nonumber\\
    &+\frac{1}{288}\left(\frac{V'''_{0}V^{(5)}_{0}}{V''^{2}_{0}}\right)(19+28\alpha^{2})-\frac{1}{288}\left(\frac{V^{(6)}_{0}}{V''_{0}}\right )(5+4\alpha^{2})\Big \}\label{12}
\end{align}
and
\begin{equation}\label{13}
    \alpha=n+\frac{1}{2},\> V^{(n)}_{0}=\frac{d^{n}V}{dr^{n}_{*}}\Big|_{r_{*}=r_{*}(r_{p})}
\end{equation}
Take $M=1, c=0.001$ and $M=1, c=0$ for our calculation. And $c=0$
 means there is no quintessence. Using the third-order WKB approximation,
 we can get the solutions as shown in the table 1 and table 2 . Where $l$ is the angular harmonic index,
 $n$ is the overtone number, $\omega$ is the complex quasinormal frequencie, $w_{q}$ is the quintessential state
parameter.\\

TABLE I: The quasinormal frequencies of gravitational perturbation
in a Schwarzshild black hole without quintessence(c=0).
\begin{equation}\label{121}
        \begin{tabular}{cccccc}
    \hline\hline
    $l$ & $n$ & $\omega$&$l$ & $n$ & $\omega$ \\
    \hline
    $2$&$0$&$0.37316-0.08922i$&$4$&$0$&$0.80910-0.09417i$\\
    $$&$1$&$0.34602-0.27492i$&$$&$1$&$0.79650-0.28437i$\\
    $$&$2$&$0.30293-0.47106i$&$$&$2$&$0.77364-0.47897i$\\
    $$&$3$&$0.24746-0.67290i$&$$&$3$&$0.74331-0.67830i$\\
    $$&$$&$$&$$&$4$&$0.70721-0.88127i$\\
    $3$&$0$&$0.59927-0.09273i$&$5$&$0$&$1.01225-0.09487i$\\
    $$&$1$&$0.58235-0.28141i$&$$&$1$&$1.00215-0.28583i$\\
    $$&$2$&$0.55320-0.47668i$&$$&$2$&$0.98326-0.47990i$\\
    $$&$3$&$0.51575-0.67743i$&$$&$3$&$0.95748-0.67780i$\\
    $$&$4$&$0.47107-0.88154i$&$$&$4$&$0.92636-0.87919i$\\
    \hline\hline
  \end{tabular}\nonumber
\end{equation}

\section{Discussion and Conclusion}
The data of table I is obtained by using the WKB method in a
Schwarzschild black hole without quintessence, and table II is under
the quintessence. Explicitly, we plot the relationship between the
real and imaginary parts of quasinormal frequencies with the
variation of $w_{q}$(for fixed $c=0.001$), and compared with no
quintessence. From the figure I we can find that for fixed
$c$(unequal to 0) and $l$ the absolute value of the real and
imaginary parts decrease as the quintessence state parameter $w_{q}$
decreases. The quasi-normal modes of the black hole present complex
frequencies whose real part represents the actual frequency of the
oscillation and the imaginary part representing the damping. It
means that when the value of $w_{q}$ is smaller, the oscillations
damps more slowly. Also the absolute value of the real and imaginary
parts of quasinormal modes with quintessence is smaller compared
with no quintessence for given $l$ and $n$. That is to say, due to
the presence of quintessence, the oscillations of the gravitational
wave damps more slowly, and that is different from the situation of
the scalar field given by Songbai Chen et al\cite{45}.

As we know, lots of stars will end their lives with a violent
supernova explosion. This will leave behind a compact object
oscillated violently in the first few seconds, emit a large amount
of gravitational radiation and the initial oscillations will damp
out. We could observe for a few tenths of a millisecond continuous
monochromatic blasting if the supernova remnants formed a black
hole. And the gravitational waves will carry away information about
the black hole. From the calculation, the paper shows that the
quintessence will influence the quasinormal modes of gravitational
perturbation around a black hole. We set $c=0.001$. $c$ is the
normalization factor related to the density of quintessence around
the black hole. Actually, $c$ may be much smaller than $0.001$ and
hence the influence to gravitational perturbation could be
neglected. Although the QNMs are predicted by the perturbation
equations, it is not always clear which ones will be excited and
under what initial conditions. Only when the density of the
quintessence surrounding the black hole is high enough to make the
influence on the QNMs distinct and measurable could we investigate
the character of quintessence by way of extraction and analysing the
experimental data.

 \setcounter{secnumdepth}{-1}

\acknowledgements{Yu Zhang wishes to thank Professor H Y Liu and Dr
LiXin Xu for helpful discussions. This work is supported by the
National Natural Science Foundation of China under Grant No.
10573004.}
\\
\\
\\

TABlE II:The quasinormal frequencies of gravitational perturbation
in a Schwarzshild black hole surrounded by quintessence for $l=2$,
$l=3$, $l=4$, $l=5$ and $c=0.001$.

 \begin{equation}\label{121}
   \begin{tabular}{cccccc}
     \hline\hline
    $3w_{q}+1$ & $\omega(n=0)$ & $\omega(n=1)$ & $\omega(n=2)$& $\omega(n=3)$ & $\omega(n=4)$\\
     \hline
    $l=2$\\
     0&$0.37261-0.08904i$&$0.34553-0.27437i$&$0.30256-0.47013i$&$0.24722-0.67155i$\\
     -0.4&$0.37230-0.08894i$&$0.34527-0.27406i$&$0.30235-0.46958i$&$0.24710-0.67078i$\\
     -0.8&$0.37182-0.08880i$&$0.34485-0.27362i$&$0.30205-0.46882i$&$0.24693-0.66967i$\\
     -1.2&$0.37107-0.08862i$&$0.34423-0.27303i$&$0.30161-0.46776i$&$0.24672-0.66810i$\\
     -1.6&$0.36990-0.08840i$&$0.34333-0.27227i$&$0.30105-0.46628i$&$0.24651-0.66585i$\\
     -2.0&$0.36810-0.08819i$&$0.34209-0.27131i$&$0.30039-0.46415i$&$0.24629-0.66240i$\\
    $l=3$\\
     0&$0.59837-0.09255i$&$0.58151-0.28085i$&$0.55242-0.47574i$&$0.51506-0.67608i$&$0.47048-0.87978i$\\
     -0.4&$0.59788-0.09244i$&$0.58104-0.28052i$&$0.55199-0.47518i$&$0.51468-0.67528i$&$0.47017-0.87874i$\\
     -0.8&$0.59712-0.09229i$&$0.58031-0.28006i$&$0.55134-0.47438i$&$0.51410-0.67415i$&$0.46968-0.87726i$\\
     -1.2&$0.59592-0.09208i$&$0.57919-0.27942i$&$0.55033-0.47329i$&$0.51324-0.67257i$&$0.46899-0.87518i$\\
     -1.6&$0.59406-0.09183i$&$0.57748-0.27862i$&$0.54887-0.47184i$&$0.51204-0.67039i$&$0.46807-0.87223i$\\
     -2.0&$0.59117-0.09156i$&$0.57495-0.27767i$&$0.54684-0.46992i$&$0.51049-0.66729i$&$0.46690-0.86787i$\\
    $l=4$\\
     0&$0.80789-0.09398i$&$0.79532-0.28380i$&$0.77252-0.47802i$&$0.74226-0.67694i$&$0.70625-0.87950i$ \\
     -0.4&$0.80723-0.09387i$&$0.79468-0.28347i$&$0.77190-0.47745i$&$0.74169-0.67614i$&$0.70572-0.87846i$ \\
     -0.8&$0.80619-0.09372i$&$0.79367-0.28299i$&$0.77094-0.47664i$&$0.74079-0.67499i$&$0.70490-0.87696i$ \\
     -1.2&$0.80457-0.09350i$&$0.79210-0.28234i$&$0.76947-0.47554i$&$0.73943-0.67340i$&$0.70367-0.87487i$ \\
     -1.6&$0.80206-0.09324i$&$0.78970-0.28151i$&$0.76726-0.47409i$&$0.73746-0.67126i$&$0.70195-0.87199i$ \\
     -2.0&$0.79815-0.09294i$&$0.78606-0.28055i$&$0.76406-0.47228i$&$0.73474-0.66840i$&$0.69966-0.86795i$ \\
    $l=5$\\
     0&$1.01074-0.09468i$&$1.00066-0.28526i$&$0.98182-0.47894i$&$0.95610-0.67644i$&$0.92505-0.87743i$\\
     -0.4&$1.00990-0.09457i$&$0.99984-0.28492i$&$0.98103-0.47837i$&$0.95534-0.67563i$&$0.92433-0.87638i$\\
     -0.8&$1.00861-0.09441i$&$0.99856-0.28444i$&$0.97979-0.47756i$&$0.95415-0.67448i$&$0.92321-0.87488i$\\
     -1.2&$1.00658-0.09419i$&$0.99658-0.28378i$&$0.97788-0.47644i$&$0.95234-0.67289i$&$0.92151-0.87280i$\\
     -1.6&$1.00344-0.09392i$&$0.99353-0.28294i$&$0.97499-0.47499i$&$0.94966-0.67078i$&$0.91906-0.86999i$\\
     -2.0&$0.99854-0.09362i$&$0.98884-0.28198i$&$0.97068-0.47324i$&$0.94581-0.66808i$&$0.91567-0.86619i$\\
    \hline\hline
   \end{tabular}\nonumber
 \end{equation}

   \begin{figure}
    \includegraphics[angle=0, width=0.6\textwidth]{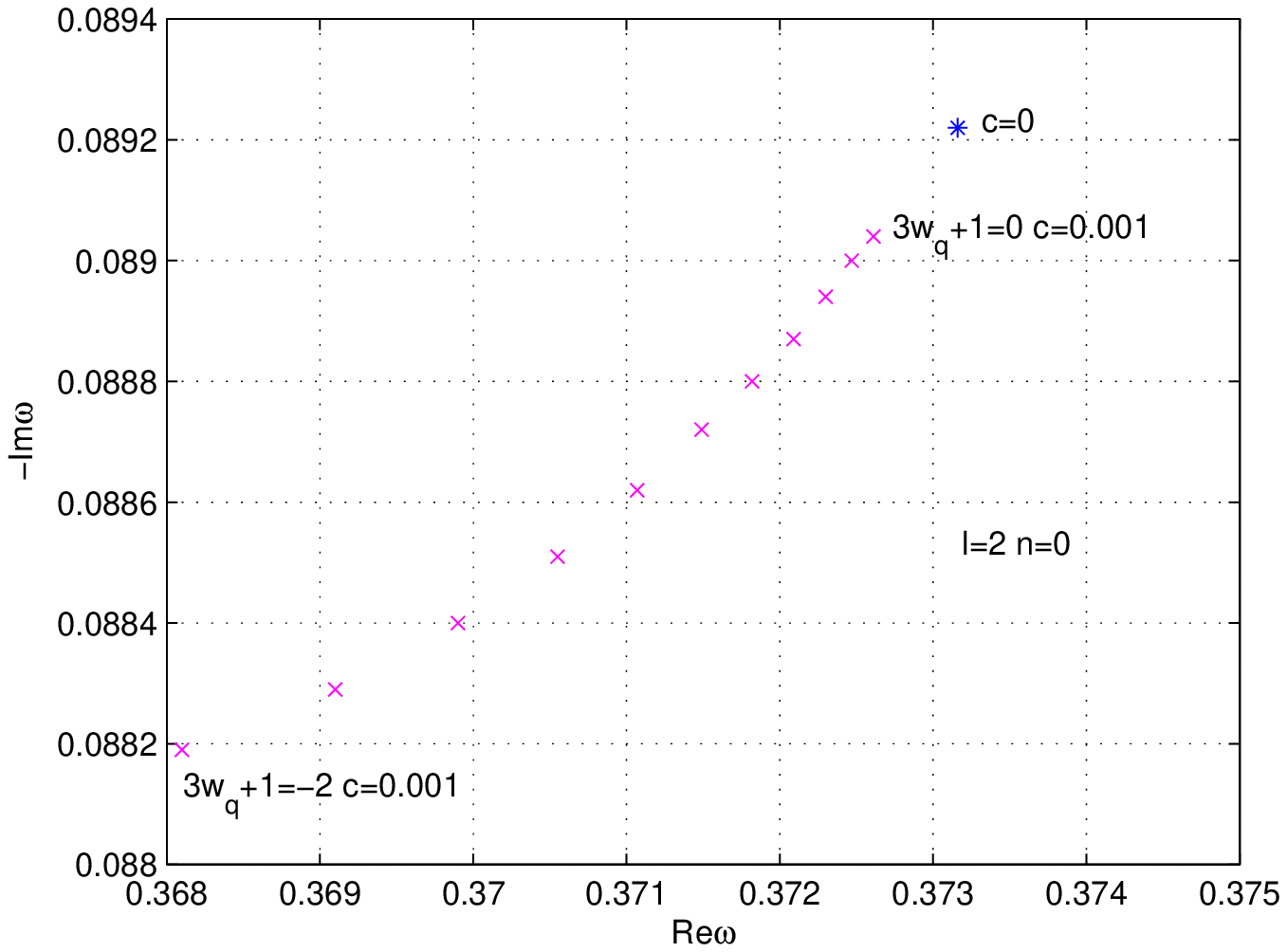}
    \includegraphics[angle=0, width=0.6\textwidth]{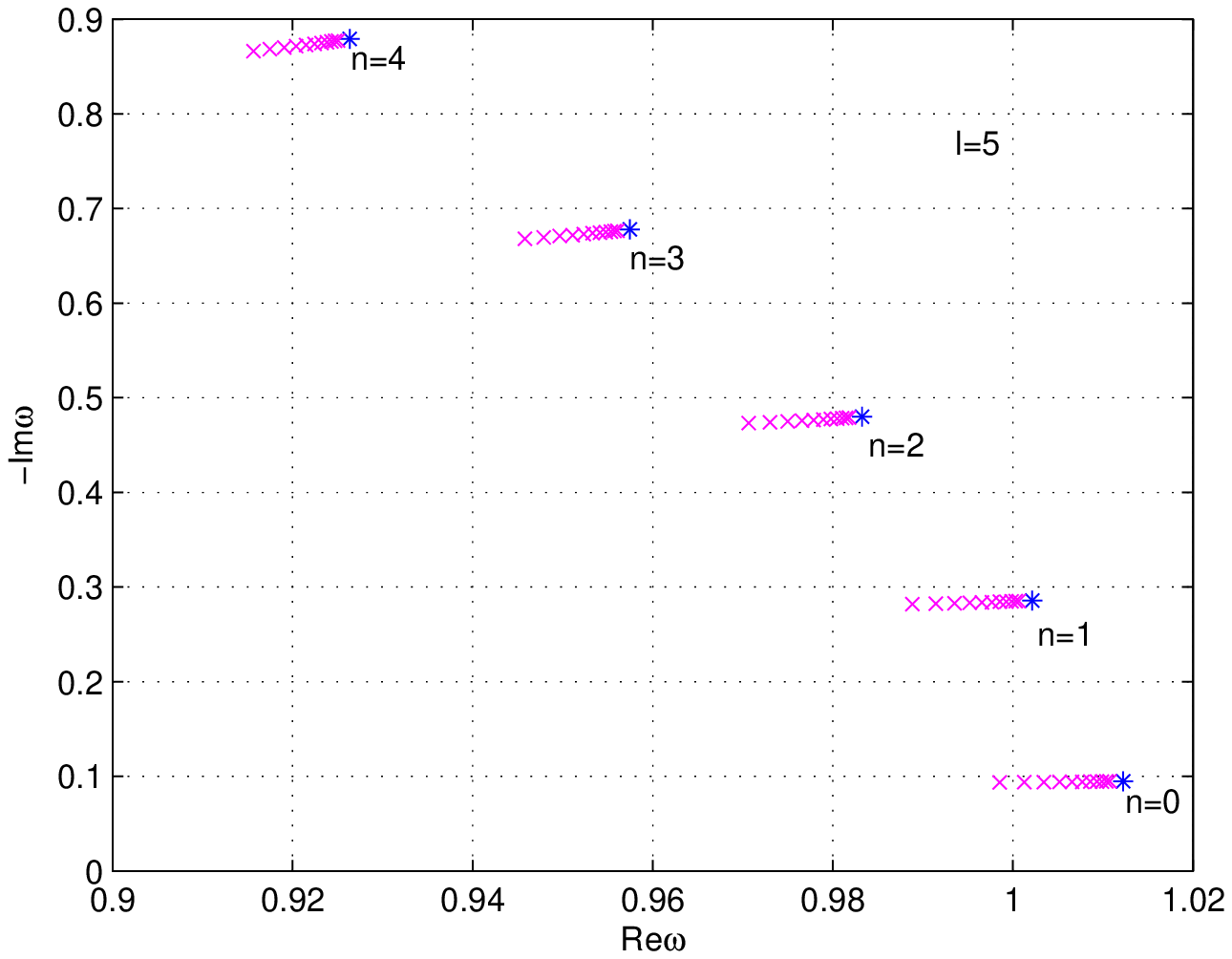}
\caption{The relationship between the real and imaginary
 parts of quasinormal frequencies of the gravitational perturbation
 in the background of the black hole surrounded by quintessence for
 fixed $c=0.001$ and no quintessence for $c=0$. $\times$ and
 $\ast$ refer to that there is quintessence and no quintessence, respectively.}
\end{figure}


\begin{thebibliography}{99}
 {\small
 \bibitem{1}C. V. Vishveshwara, Nature {\bf 227} 936 (1970).
 \bibitem{2}S. Chandrasekhar, Proc. R. Soc. London, Ser. A. {\bf 343}, 289-298 (1975).;  S. Chandrasekhar, and S. Detweiler,
 Proc. R. Soc. London, Ser. A. {\bf344}, 441-452 (1975).
\bibitem{3}H. P. Nollert, Phys. Rev. D {\bf 47}, 5253-5258 (1993).
\bibitem{4}E. W. Leaver, Proc. R. Soc. A {\bf 402} 285 (1985) .
\bibitem{5}K. D. Kokkotas and B. G. Schmidt, Living Rev. Rel. {\bf 2}2 (1999); H. P. Nollert, Class. Quant. Grav. {\bf 16}, R159 (1999).
\bibitem{6}V. Cardoso and J. P. S. Lemos£¬ Phys. Rev. D 64, 084017 (2001);
E. Beri, V. Cardoso and S. Yoshida, Phys. Rev. D{69} 124018 (2004);
  V. Cardoso, J. P. S. Lemos, S. Yoshida, Phys. Rev. D{\bf 70} 124032 (2004) .
\bibitem{7}A. J. M. Medved, D. Martin, M. Visser, Class. Quant. Grav. {\bf 21} 2393-2405 (2004) .
 A. J. M. Medved, D. Martin, Gen. Rel. Grav. {\bf 37} 1529-1539 (2005)
\bibitem{8}H. B. Zhang, Z. J. Cao, X. F. Gong, W. Zhou,Class. Quant. Grav. {\bf 21} 917-925 (2004)
\bibitem{9}J. L. Jing, Phys. Rev. D{\bf 69} 084009 (2004) ;
S. B. Chen, J. L. Jing, Class. Quant. Grav. {\bf 22} 533-539 (2005)
; S. B. Chen, J. L. Jing, Class. Quant. Grav. {\bf 22} 2159-2165
(2005) .
\bibitem{10}R. Konoplya, Phys. Rev. D{\bf 71} 024038 (2005).
\bibitem{11}I. B. Khriplovich, Int. J. Mod. Phys. D{\bf 14} 181 (2005).
\bibitem{12}F. W. Shu, Y.G. Shen, Phys. Rev. D{\bf 70} 084046 (2004).
\bibitem{13}J. Cris\'{o}stomo, S. Lepe, J. Saavedra, Class. Quant. Grav. {\bf 21}
2801-2809 (2004).
\bibitem{14}F. W. Shu, Y. G. Shen, Phys. Lett. B{\bf 614}
195-200 (2005).
\bibitem{15}R. G. Daghigh, G. Kunstatter, Class. Quant. Grav. {\bf 22}
4113-4128 (2005).
\bibitem{16}E. Berti, K. D. Kokkotas, Phys. Rev. D{\bf 71}
124008 (2005).
\bibitem{17}M. Giammatteo, I. G. Moss, Class. Quant. Grav. {\bf 22}
1803-1824 (2005).
\bibitem{18}B. F. Schutz, and C. M. Will,  Astrophys. J.{\bf 291}, L33-L36 (1985).
\bibitem{19}S. Iyer and C. M. Will, Phys. Rev. D {\bf 35}, 3621 (1987)
\bibitem{20}R. A. Konoplya, Phys. Rev. D {\bf 68} 024018 (2003)
\bibitem{21}S. Iyer, Phys. Rev. D {\bf 35}, 3632 (1987);
\bibitem{22}K. D. Kokkotas and B. F. Schutz, Phys. Rev. D {\bf 37}, 3378 (1988)
\bibitem{23}E. Seidel, and S. Iyer, Phys. Rev. D {\bf 41}, 374-382 (1990).
\bibitem{24}K. D. Kokkotas, Nuovo Cimento B {\bf 108}, 991-998 (1993).
\bibitem{25} J. M. Maldacena, Adv. Theor. Math. Phys. {\bf 2}, 231 {1998}.
\bibitem{26} E. Witten, Adv. Theor. Math. Phys. {\bf 2}, 253 {1998}.
\bibitem{27} Gubser S S Phys. Rev. D{\bf 63} 084017  {2001}
\bibitem{28}B. Wang, C. Y. Lin and E. Abdalla, Phys. Lett. B {\bf 481}, 79 (2000). B. Wang, E. Abdalla and R. B.
Mann, Phys. Rev. D {\bf 65}, 084006 (2002); B. Wang, C.Y. Lin and C.
Molina, Phys. Rev. D {\bf 70}, 064025 (2004).
\bibitem{29}V. Cardoso and J. P. S. Lemos, Phys. Rev. D {\bf 64}, 084017 (2001); V. Cardoso and J. P. S. Lemos, Class.
Quantum. Grav. {\bf 18}, 5257 (2001).
\bibitem{30}D. Birmingham, I. Sachs, S. N. Solodukhin, Phys. Rev. Lett. {\bf 88}, 151301 (2002); D. Birmingham, Phys.
Rev. D {\bf 64}, 064024 (2001).
\bibitem{31}R. A. Konoplya, Phys. Rev. D {\bf 66}, 044009 (2002).}
\bibitem{32}S. Hod, Phy. Rev. Lett. {\bf 81}, 4293 (1998).
\bibitem{33}A. Corichi, Phys. Rev. D {\bf 67}, 087502 (2003).
\bibitem{34}O. Dreyer, Phy. Rev. Lett. {\bf 90}, 081301 (2003).
\bibitem{35}A. D. Miller et al., Astrophys. J. Lett. {\bf  524}, L1 (1999);
 P. de Bernardis et al., Nature {\bf 404}, 955 (2000);
 S. Hanany et al., Astrophys. J. Lett. {\bf 545}, L5 (2000);
 N. W. Halverson et al., Astrophys. J. {\bf 568}, 38 (2002);
 L. Page et al., Astrophys. J. Suppl. {\bf 148}, 233 (2003);
 S. Hannestad, G. Raffelt, Phys. Rev. D {\bf 72} 103514 (2005);
 R. Lamon, R. Durrer,  Phys. Rev. D {\bf 73} 023507 (2006).
\bibitem{36}P. Ruiz-Lapuente, A. Burkert, R. Canal, Astrophys. J. {\bf 447} L69
(1995); A. G. Riess et al., Astron. J. {\bf 116}, 1009 (1998);
 D. Branch, Ann. Rev. Astron.Astrophys. {\bf 36} 17-55 (1998);
 R. A. Knop et al., Astrophys. J. {\bf 598}, 102 (2003).
 A. G. Riess et al., Astrophys. J. {\bf 607}, 665 (2004);
 X. Zhang, F. Q. Wu, Phys. Rev. D {\bf 72}  043524 (2005).
\bibitem{37}D. J. Bacon et al., MNRAS, {\bf 318}, 625 (2000).
D. J. Bacon et al., MNRAS, {\bf 344}, 673 (2003). R. Scranton et
al., astro-ph/0307335; N. Kaiser et al., astro-ph/0003338;
 M. Tegmark et al., Phys. Rev. D {\bf 69}, 103501 (2004).
\bibitem{38}T. Kitayama, Y. Suto, Astrophys. J. {\bf 469} 480 (1996)
 ; J. V. Cunha, J. S. Alcaniz, J. A. S. Lima, Phys. Rev. D{\bf 69}
083501 (2004); Y. Wang, M. Tegmark, Phys. Rev. Lett. {\bf 92} 241302
(2004) .
\bibitem{39}S. Weinberg, Rev. Mod. Phys. {\bf 61}, 1 (1989);
K. A. Olive, M. Pospelov, Phys. Rev. D {\bf 65} 085044 (2002); J. S.
Alcaniz, Phys. Rev. D {\bf 69} 083521 (2004).
\bibitem{40}B. Ratra and P. J. E. Peebles, Phys. Rev. D {\bf 37}, 3406 (1988);
 P. J. E. Peebles and B. Ratra, Astrophys. J. {\bf 325}, L17 (1988);
 R. RCaldwell, R. Dave, and P. J. Steinhardt, Phys. Rev. Lett. {\bf 80}, 1582
(1998);
 I. Zlatev, L. Wang and P. J. Steinhardt, Phys. Rev. Lett. {\bf 82},
896 (1999); V. Sahni, L. M. Wang, Phys. Rev. D {\bf 63}, 103517
(2000); T. Matos, L. A. Urena-Lopez , Phys. Rev. D {\bf 63}, 063506
(2001); S. Capozziello, V.F. Cardone, E. Piedipalumbo, C. Rubano,
Class.Quant.Grav. {\bf 23}, 1205-1216 (2006).
\bibitem{41} A. E. Schulz, M. White, Phys. Rev. D {\bf 64}, 043514 (2001);
R. R. Caldwell, Phys. Lett. B {\bf 545}, 23 (2002); S. Nojiri and S.
D. Odintsov, Phys. Lett. B {\bf 562}, 147 (2003); S. Nojiri and S.
D. Odintsov, Phys. Lett. B {\bf 565}, 1 (2003); P. Singh, M. Sami,
and N. Dadhich, Phys. Rev. D {\bf 68}, 023522 (2003);  L. P.
Chimento and R. Lazkoz, Phys. Rev. Lett. {\bf 91}, 211301 (2003); A.
Vikman, Phys. Rev. D {\bf 71},  023515 (2005).
\bibitem{42} T. Chiba, T. Okabe and M. Yamaguchi, Phys. Rev. D {\bf 62}, 023511
(2000); R. J. Scherrer, Phys. Rev. Lett. {\bf 93}, 011301 (2004); L.
P. Chimento, M. Forte, R. Lazkoz, Mod. Phys. Lett. A {\bf 20} 2075
(2005).
\bibitem{43}B. Feng, M. Z. Li, Y. S. Piao, X. M. Zhang, Phys. Lett. B {\bf 634} 101-105 (2006)
; H. Wei, R. G. Cai, D. F. Zeng, Class. Quant. Grav. {\bf 22}
3189-3202 (2005); G. B. Zhao, J. Q. Xia, M. Z. Li, B. Feng, X. M.
Zhang, Phys. Rev. D {\bf 72} 123515(2005).


\bibitem{44}V. V. Kiselev, Class. Quant. Grav. {\bf
20}, 1187-1197 (2003).
\bibitem{45}S. B. Chen, J. L. Jing, Class. Quant. Grav. {\bf 22},
4651-4657 (2005).
\bibitem{46}T. Regge , J. A. Wheeler, Phys. Rev., V. {\bf
108}, 1063-1069 (1957).
\bibitem{47}F. J. Zerilli, Phys. Rev. Lett. {\bf 24}, 737 (1970).
\bibitem{48}L. P. Einsenhart, Riemannian Geometry, Princeton Univ,
press, Princeton (1926), Chap. VI.

\end{thebibliography}
\end{document}